\documentclass[reprint]{JASA}
\newcommand{\R}{\mathbb{R}}
\newcommand{\N}{\mathbb{N}}

\def\app{\text{app}}

\newcommand{\dd}{\mathrm{d}}

\newcommand{\E}{\mathbb{E}}

\newcommand{\CV}{\text{Cov}}

\newcommand\tint{\textstyle\int\nolimits}

\usepackage{amsthm} 
\theoremstyle{definition}
\newtheorem{defi}{Definition}

\newtheorem{prop}{Proposition}

\usepackage{color}
\usepackage{colortbl}

\begin{document}

\title[Automated approach for source location in shallow waters]{Automated approach for source location in shallow waters}
\author{Angèle Niclas}
\email{angele.niclas@polytechnique.edu}
\author{Josselin Garnier}
\email{josselin.garnier@polytechnique.edu}
\affiliation{CMAP, CNRS, École polytechnique, Institut Polytechnique de Paris,  Palaiseau, 91120, France}

\preprint{Author, JASA}		

\date{\today} 

\begin{abstract}
This paper proposes a fully automated method for recovering the location of a source and medium parameters in shallow waters. The scenario involves an unknown source emitting low-frequency sound waves in a shallow water environment, and a single hydrophone recording the signal. Firstly, theoretical tools are introduced to understand the robustness of the warping method and to propose and analyze an automated way to separate the modal components of the recorded signal. Secondly, using the spectrogram of each modal component, the paper investigates the best way to recover the modal travel times and provides stability estimates. Finally, a penalized minimization algorithm is presented to recover estimates of the source location and medium parameters. 
The proposed method is tested on experimental data of right whale gunshot and combustive sound sources, demonstrating its effectiveness in real-world scenarios.   
\end{abstract}


\maketitle

\section{\label{sec:1} Introduction}
This paper presents a fully automated method for recovering the location of a source and medium parameters in shallow waters. Specifically, we consider a scenario where an unknown source emits low-frequency sound waves (typically less than 500 Hz) in a coastal environment, and a single hydrophone records the signal at a distance of more than 1 km from the source. The ability to accurately recover the source location from the recorded signal has important military applications for localizing quiet sources~\cite{ainslie1}, as well as for monitoring fish~\cite{stanton1} and whale populations~\cite{mellinger1}. 

To study this inverse problem and propose a  fully automated resolution method, we model the source as a stationary point source and the shallow water environment as a semi-infinite Pekeris waveguide \cite{pekeris1} with a finite layer of water and an infinite layer of sediments. 
Under these ideal circumstances
the Fourier transform $\hat{u}$ of the recorded signal $u$ can be decomposed into modal components,
\begin{equation}
\label{eq:decomp}
\hat{u}(\omega)=\sum_{n\in \N} A_n(\omega)e^{i\Phi_n(\omega)},
\end{equation}
where $n$ is the mode index, $A_n$ is the complex amplitude of mode $n$, $\Phi_n=-k_n r$ with $k_n$ the horizontal wavenumber of mode $n$ and $r$ the distance from the source to hydrophone \cite{jensen1}. 
However, due to uncertainties in the propagation environment and the source characteristics, the modal decomposition (\ref{eq:decomp}) of the recorded signal $\hat{u}$ is not exact and is difficult to extract and process.

Different techniques to extract information about the source and environment have been proposed.
One of the most commonly used techniques is called Time-of-Arrival (TOA) estimation (see for instance \cite{bonnel3,aubauer1}). This method involves measuring the modal travel times $t_n(\omega)=-\Phi_n'(\omega)$ (where the prime stands for the frequency derivative) at the hydrophone and using this information to estimate the distance to the source. Methods have been developed to retrieve directly the phases $\Phi_n$ from the signal $\hat{u}$ \cite{bonnel4}, but there are not very effective when data are noisy. Most existing methods take advantage of the modal dispersion curves in the time-frequency domain, and consists in warping the signal to extract each modal component, recovering the dispersion curves $\omega\mapsto t_n(\omega)$ associated with each mode using time-frequency analysis, and matching estimated dispersion curves with simulated replicas. 

The warping method is a widely used technique that aims at increasing the distance between modes in the time-frequency space, making it easier to separate them (see a full description in \cite{bonnel1}). Its main advantage is its robustness to environmental uncertainties. However, as pointed out in \cite{bonnel1}, "Warping is not a universal solution and requires expertise and judgment to be used effectively, as there are currently no automated methods for bulk warping." This comment also applies to the recovery of dispersion curves using time-frequency analysis, where there is no known automated method for recovering dispersion curves with stability estimates. Most experimental data are thus processed manually, depending on the situation \cite{bonnel2,thode1}. The inverse problem, usually presented as a minimization problem between the estimated dispersion curves and simulated replicas, also lacks of an automated procedure. In practice, the minimization problem is solved numerically by computing the function's value everywhere and taking its minimum value, which can be computationally expensive. 

This paper proposes a solution to these limitations by presenting a completely automated method to recover source location in shallow waters. The design of this automated method is based on the warping method \cite{bonnel2} and on several additional theoretical arguments. Specifically, we provide theoretical tools to understand the robustness of the warping method and use a watershed algorithm to implement it automatically. We then present two methods to recover the dispersion curves $\omega\mapsto t_n(\omega)$ and quantify their stability with noisy data. Finally, we implement a penalized minimization algorithm that can use prior estimates on some medium parameters (if available) to estimate the source location. 

The paper is organized as follows: Section~\ref{sec:general} describes the general framework of the Pekeris waveguide and the associated modal decomposition. In Section~\ref{sec:extraction}, we focus on the warping technique and provide an automated way to separate modal components in the signal. In Section~\ref{sec:reconstruction}, we introduce and compare the maximum and mean methods for recovering dispersion curves and provide stability estimates in both cases. Finally, in Section~\ref{sec:inverse}, we present a penalized minimization algorithm that makes it possible to estimate the source location and medium parameters.

\section{General framework}
\label{sec:general}%
\subsection{Acoustic waves propagation}
\begin{figure}
\includegraphics{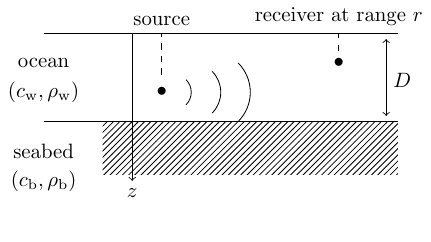}\vspace{-5mm}
\caption{\label{pekeris} Scheme of the Pekeris waveguide.}
\vspace{-0.5cm}
\end{figure}

Let us model the shallow water environment by a Pekeris waveguide $\R^2 \times (0,+\infty)$ represented in Figure \ref{pekeris} and formed with a thin layer of ocean for depths between $0$ and $D$ and an infinite layer of sediment for depths greater than $D$  \cite{pekeris1}. The celerity $c$ and the density $\rho$ are assumed to be piecewise constant, and an impulsive source 
transmits a signal received by a receiver (hydrophone)
located at range $r$. The case of non-impulsive signals can be handled similarly using deconvolution techniques or dispersion curve differences described in \cite{bonnel1}. 
The acoustic wave~$u$ emitted by the impulsive source satisfies the acoustic wave equation 
\begin{equation}
\rho \nabla \cdot \left(\frac{1}{\rho} \nabla u\right)-\frac{1}{c^2}\partial_{tt} u=0,
\end{equation}
with the free surface (Dirichlet) boundary condition $u_{|z=0}=0$. Here $\rho$ is the density and $c$ is the wave speed. 
Following the analysis presented in \cite[§2.4.5]{jensen1}, 
the Fourier transform  $\hat{u}(\omega)=\tint_{\mathbb{R}} u(t) \exp(-i\omega t) \dd t$ of the signal $u(t)$ recorded by the receiver can be decomposed as a sum of modal components
\begin{equation}
\hat{u}(\omega)=\sum_{n\in \N} \hat{u}_n
(\omega)=\sum_{n\in \N} A_n(\omega)e^{i\Phi_n(\omega)},
\end{equation}
where the complex amplitude $A_n$ is slowly varying in $\omega$ and depends
on the shape of the emitted signal and the depths and ranges of the source and receiver \cite{jensen1} and the phase is of the form $\Phi_n (\omega)=-k_n(\omega)r$, where $r$ is the distance from the source to the receiver (range) and $k_n$ is the $n$-th wavenumber given by the $n$-th solution of the dispersion relation
\begin{equation}\label{disp}
\tan\left(D\sqrt{\omega^2/c_{\rm w}^2}\right)=-\frac{\rho_{\rm b}}{\rho_{\rm w}}\sqrt{\frac{\omega^2/c_{\rm w}^2-k_n^2}{k_n^2-\omega^2/c_{\rm b}^2}}.
\end{equation}
Given the little information that we have, the general idea is to use $\Phi_n(\omega)$ and \eqref{disp} to recover information about $r$ and the parameters of the Pekeris waveguide. However, phrase retrieval is difficult with noisy data (see however \cite{bonnel4}), and it is easier to retrieve the modal travel time $t_n(\omega)$ defined by 
\begin{equation}
t_n(\omega)=-\Phi_n'(\omega)=r k_n'(\omega).
\end{equation}

\begin{figure}
\begin{tabular}{ll}
\includegraphics[width=0.23\textwidth]{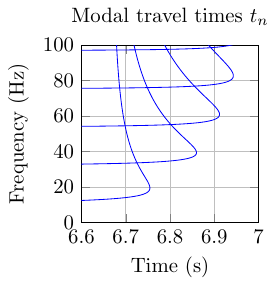} &\includegraphics[width=0.23\textwidth]{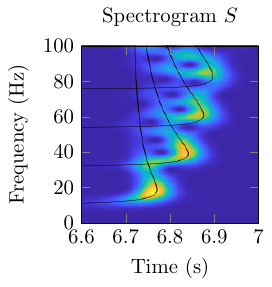} \vspace{-0.5cm}\\ 
(a) & (b)\end{tabular}
\caption{\label{spectro} (a): Example of dispersion curves $\omega \mapsto t_n(\omega)$ in the Pekeris model with parameters given in Table \ref{tableau}-line 1. (b): Spectrogram $S$ computed with a Gaussian window defined in \eqref{h} with $\sigma=20$ Hz, superposed with the dispersion curves. 
}
\end{figure}

In Figure \ref{spectro}, we plot an example of the dispersion curves $\omega \mapsto t_n(\omega)$. The time-frequency representation \cite{boashash1} of signals is known to concentrate energy around these dispersion curves, which enables to recover $r$ and the parameters of the Pekeris waveguide. Following the investigation led in~\cite{bonnel3} on the most appropriate time-frequency representation, we choose to use the short-time Fourier transform (STFT) of the signal and the spectrogram $S$ defined by 
\begin{align}S(t,\omega)&=|STFT(t,\omega)|^2, \\
STFT(t,\omega)&=\int_\R u(\tau)h(\tau-t)e^{-i\omega \tau}\dd \tau\\ 
& =\frac{e^{i\omega t}}{2\pi}\int_\R \hat{u}(\xi+\omega)\hat{h}(\xi)e^{i\xi t}\dd\xi.
\end{align}
Similarly, we define the modal component $S_n$ of the spectrogram $S$ by 
\begin{align}\label{def_spectro}
S_n(t,\omega)&=\left|\int_\R u_n(\tau)h(\tau-t)e^{-i\omega \tau}\dd \tau\right|^2\\ &=\frac{1}{4\pi^2}\left|\int_\R \hat{u}_n(\xi+\omega)\hat{h}(\xi)e^{i\xi t}\dd\xi\right|^2. 
\end{align}

In this definition, $h$ is a centered window function. Different choices of $h$ are possible (rectangular windows, sine windows, Hamming windows...), and in the following, we choose to work with a Gaussian window of standard deviation $\sigma$ to simplify the computations. The function $h$ is then defined by 
\begin{equation}\label{h}
h(t)=\frac{\sigma}{\sqrt{2\pi}}
\exp\Big( -\frac{\sigma^2 t^2}{2}\Big), \quad \hat{h}(\omega)=\exp\Big( -\frac{\omega^2}{2\sigma^2} \Big).
\end{equation}

An example of a spectrogram with $\sigma=20$ Hz is presented in Figure \ref{pekeris}.
As explained in \cite{bonnel1}, if $\sigma$ is small enough, the support of $\hat{h}$ is small and we can hope to simplify the expression of $S$. The general idea, detailed in the following sections, first uses the fact that modes can be separated and that one can use $S$ to recover the modal spectrograms $S_n$ that are of the form
\begin{equation}
S_n(t,\omega)\!=\!\frac{1}{4\pi^2}\left|\int_{\R} \!\!A_n(\xi+\omega)e^{i\Phi_n(\xi+\omega)}\hat{h}(\xi)e^{i\xi t}\dd \xi\right|^2. 
\end{equation}
Then, using the fact that the support of $\hat{h}$ is small,
\begin{align}\label{approx}
A_n(\xi+\omega)&\approx A_n(\omega), \\ \Phi_n(\xi+\omega)&\approx \Phi_n(\omega)+\xi\Phi_n'(\omega)=\Phi_n(\omega)-\xi t_n(\omega),
\end{align}
and we can find a direct link between $S_n(t,\omega)$ and $t_n(\omega)$:
\begin{align}\label{sapp}
S_n(t,\omega)&\approx \frac{1}{4\pi^2}|A_n(\omega)|^2 \left| \int_\R \hat{h} (\xi) e^{i\xi(t-t_n(\omega))}\dd \xi\right|^2\\ & = |A_n(\omega)|^2 h(t-t_n(\omega))^2.
\end{align}
We will detail these approximations in the following, as well as the use of spectrograms to recover the modal travel times and the environment parameters. Throughout the paper, we will illustrate our method on synthetically generated data and on experimental data presented in the following subsection.

\subsection{Synthetical and experimental data}

In this paper, all numerical simulations on synthetically generated data are presented for a Pekeris model whose parameters are given in the first line of Table \ref{tableau}. These data are generated on Matlab using the code developed by J. Bonnel \cite{bonnel1} where the Fourier transform $\hat{u}$ is computed using the modal expansion presented in~\cite{jensen1}, and then moved into the time domain using an inverse fast Fourier transform. Spectrograms are calculated using the time-frequency toolbox \cite{tft1}, and enable to recover four different propagative modes in the modal decomposition. 

We also implement our algorithm on experimental data from two different experiments. The first one is the recording of a right whale gunshot presented in \cite{thode1}, where four propagative modes are recorded with a maximal frequency $f_{\max}=200$ Hz. In this environment, we have some prior estimates of parameters given in the second line of Table \ref{tableau}. 
The second one is the recording of a combustive sound source (CSS) presented in \cite{bonnel2}, where 11 modes are clearly visible and 7 additional modes are hardly visible. In this environment, we have some prior estimates of parameters given in the third line of Table~\ref{tableau}. 

\begin{table*}\centering \scalebox{0.9}{
\begin{tabular}{|c|c|c|c|c|c|c|c|}\hline 
& $r$ (km) & $D$ (m) & $c_{\rm w}$ (m/s)  & $c_{\rm b}$ (m/s) & $\rho_{\rm w}$ (kg/m\textsuperscript{3}) & $\rho_{\rm b}$ (kg/m\textsuperscript{3}) & $f_{\max}$ (Hz) \\ \hline 
Pekeris model & 10 & 100 & 1500 & 1600 & 1000 & 1500 & 100 \\ \hline 
Right whale \cite{thode1} & 8.8 & 51 & 1450 &  ? & 1000 & 1600 & 204 \\ \hline 
CSS \cite{bonnel2} & 4.8 & 69.5 & 1464.5 & ? &  1000 & 1600 & 488 \\ \hline 
\end{tabular}}
\caption{\label{tableau}Numerical values for parameters in the Pekeris waveguide. Values are exact for the Pekeris model (line 1) and estimated for the right whale gunshot (line 2) and the combustive sound source (line 3). }\vspace{-0.5cm}
\end{table*}

\section{Extraction of the modal components of the spectrogram}
\label{sec:extraction}%
This section aims to extract the modal components $S_n$ of the spectrogram from the full spectrogram~$S$. We use the warping method described in \cite{bonnel1}. The general idea of this method is to apply a change of variable in the time axis to improve the separation of the modal components in the spectrogram. We first provide some theoretical tools to assess the mode separability. Then, we describe a way to implement this method numerically in a completely automated way.

\subsection{Warping method}

Let us denote $\omega\mapsto t_n(\omega)$ the modal travel time and $t\mapsto \omega_n(t)$ its reciprocal function. Given the fact that most of the energy of the Gaussian window $h$ (resp. $\hat{h}$) is contained in the interval $(-\sigma^{-1},\sigma^{-1})$ (resp. $(-\sigma,\sigma)$), the energy of each modal component of the spectrogram is concentrated in the rectangle
\begin{multline}
(\min(t_n)-\sigma^{-1},\max(t_n)+\sigma^{-1}) \\ \times (\min(\omega_n)-\sigma, \max(\omega_n)+\sigma). 
\end{multline}
If the intersection of these rectangles for different modes is empty, we say that the modes are well separated and we can easily extract the different modal components of the spectrogram. However, these rectangles often intersect, making this separation impossible. Hence the use of warping methods. 

The general idea of the warping method is to notice that in a perfectly reflecting waveguide with a boundary condition $\partial_z u_{|z=D}=0$ at depth $z=D$, dispersion curves are explicit and given by
\begin{equation}
\omega_n(t)=\frac{t c_{\rm w} \pi (2n-1)}{2D\sqrt{t^2-r^2/c_{\rm w}^2}} 
\end{equation}
when the emission of the pulse is at time $0$.
To improve the separability of modal components, one can change the time variable with a change of variable $t\mapsto \psi(t)$ and consider the spectrogram of the signal

\begin{equation}
\label{eqwarp} \widetilde{u}(t)
=\sqrt{|\psi(t)|}\, u\circ \psi(t).
\end{equation}
The dispersion curves of the transformed signal are linked to the ones of the original signal through the relation 
$\widetilde{\omega}_n(t)=\psi'(t)\omega_n(\psi(t))$, which enables to compute the new dispersion curves: 
\begin{equation}
\widetilde{\omega_n}(t) =\psi'(t) \frac{\psi(t) c_{\rm w} \pi (2n-1)}{2D\sqrt{\psi(t)^2-r^2/c_{\rm w}^2}}.
\end{equation}
To simplify these new curves, choosing $\psi(t)=\sqrt{t^2+r^2/c_{\rm w}^2}$ is natural (assuming that the range $r$ is known or can be estimated). This change of variable leads to constant dispersion curves
\begin{equation}
\widetilde{\omega_n}(t)= \frac{c_{\rm w} \pi (2n-1)}{2D}
\end{equation}
which are separable using a small value of $\sigma$. 

\begin{figure*}
\begin{center}
\begin{tabular}{llllll}
\includegraphics[width=0.17\textwidth]{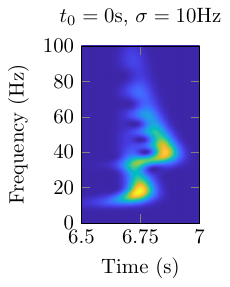} & \hspace{-4mm}
\includegraphics[width=0.17\textwidth]{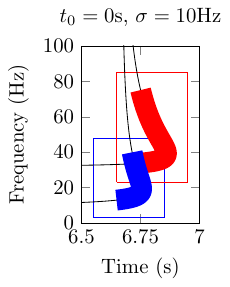} &\hspace{-4mm}
\includegraphics[width=0.17\textwidth]{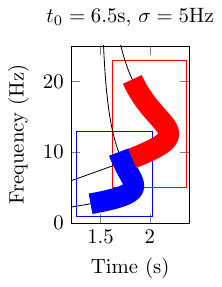} &\hspace{-4mm}
\includegraphics[width=0.17\textwidth]{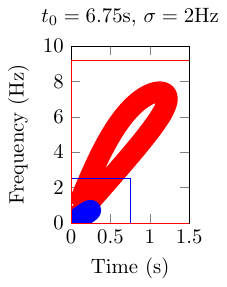} &\hspace{-4mm}
\includegraphics[width=0.17\textwidth]{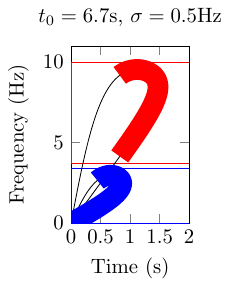} &\hspace{-4mm}
\includegraphics[width=0.17\textwidth]{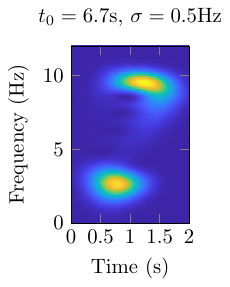} \vspace{-0.5cm}\\
(a) & (b) & (c) & (d) & (e) & (f) \end{tabular}
 \includegraphics{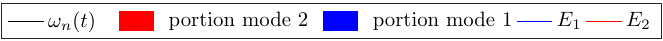} \end{center}\vspace{-0.5cm}
\caption{\label{wrap_spectro} Illustration of the warping of mode 1-2 in the Pekeris model for different choices of~$t_0$. (a) initial spectrogram computed with $\sigma=10$ Hz. (b) simplified diagram with a portion of mode 1 and 2. (c) new diagram with the warping $\psi(t)=\sqrt{t^2+t_0^2}$ with $t_0=6.5$ s. (d) new diagram with $t_0=6.75$ s. (e) new diagram with $t_0=6.7$ s. (f) new spectrogram of the warped signal at $t_0=6.7$ s, computed with $\sigma=0.5$ Hz.} \vspace{-0.5cm}\end{figure*}

In the Pekeris waveguide, the warping is more complicated. Given the similarities between Pekeris waveguides and perfectly reflecting waveguides, we choose to use the same type of warping and we define $\psi(t)=\sqrt{t^2+t_0^2}$ for a certain value $t_0>0$. The new dispersion curves are then illustrated in Figure \ref{wrap_spectro} and given by 
\begin{equation}\label{new_curves}
\widetilde{\omega_{n}}(t)=\frac{t}{\sqrt{t^2+t_0^2}}\omega_n\left(\sqrt{t^2+t_0^2}\right). 
\end{equation}

Since dispersion curves intersect, it is impossible to find a change of variable that completely separates the curves as in the perfectly reflecting waveguide. Hence, we now only aim at separating parts of dispersion curves that do not overlap (see an illustration in Figure \ref{wrap_spectro}) and contain most of the modal energy. These parts are defined as follows.

\begin{defi}\label{defi}
For each curve $\omega\mapsto t_n(\omega)$, we define $T_n^{(2)}=\max_\omega (t_n(\omega))$ and $\Omega_n^{(2)}=\text{argmax}_\omega(t_n(\omega))$ and we consider portions of the dispersion curves contained in an interval $(\Omega_n^{(1)},\Omega_n^{(3)})$ such that $\Omega_n^{(1)}<\Omega_n^{(2)}<\Omega_n^{(3)}$. We denote $T_n^{(j)}=t_n(\Omega_n^{(j)})$ for $j=1,2,3$. The energy rectangle associated to the mode number $n$ is then defined by 
\begin{multline}
E_n(\sigma):=(\min(T_n^{(1)},T_n^{(3)})-\sigma^{-1},T_n^{(2)}+\sigma^{-1})\\ \times (\Omega_n^{(1)}-\sigma,\Omega_n^{(3)}+\sigma),
\end{multline}
and we consider that two modal components $n$ and $m$ are separated if there exists $\sigma>0$ such that
\begin{equation}
 E_n(\sigma)\cap E_m(\sigma)=\varnothing. 
\end{equation}
\end{defi} 
An illustration is provided in Figure \ref{wrap_spectro}. 
Figures \ref{wrap_spectro}b show that, no matter the choice of $\sigma$, modal components cannot be separated, making warping necessary.
Figures \ref{wrap_spectro}c-d-e show the spectrogram and the dispersion curves of the warped signal.

The following proposition quantifies the separability of modal components:

\begin{prop}
\label{prop1}
Given two portions of dispersion curves associated to modes $n<m$ that do not overlap, there exists a maximum value $\Omega_{\max}$ such that if $\Omega^{(3)}_m<\Omega_{\max}$, then there exists a warping and parameters $t_0>0, \sigma>0$ that separate the two modal components $n$ and $m$. 
\end{prop}

The proof is provided in supplementary. This proposition shows that if $\Omega_m^{(3)}$ is not too large, then portions of modal dispersion curves can be separated with an appropriate choice of $t_0$ and $\sigma$. As explained in the proof, the parameter $\sigma$ needs to be chosen as large as possible in order to separate modal components in the vertical direction (see an illustration in Figure \ref{wrap_spectro}-e). Regarding the choice of $t_0$, the proof does not provide a concrete way to compute it. As illustrated in Figures \ref{wrap_spectro}d-e, a small change in $t_0$ may produce very different dispersion curves. We give in the next section a method to choose $t_0$ in an automated and appropriate way.

Let us also point out that in all the numerical and experimental examples shown below, $\Omega_{\max}$ turns out to be much larger than the maximum accessible frequencies, ensuring that $\Omega_m^{(3)}<\Omega_{\max}$. Furthermore, one can always ensure that parts of curves do not overlap by considering smaller portions of the dispersion curves and keeping only the high energy levels in the spectrogram.

\subsection{Numerical implementation}

To numerically implement the previous warping method, one must find a good value of $t_0$ and separate modes in the spectrogram. In this subsection we develop an algorithm that iteratively separates each mode, using the smallest possible value for $\sigma$ and starting with the largest propagative mode number $N$. Assuming that $t_0$ is well chosen, each modal component is separated from the others in the spectrogram. To separate them automatically, we use a watershed transform which returns a set of drainage basins associated with local maximum points of the spectrogram in the time-frequency space $(t,\omega)$. The watershed transform is an algorithm used for image processing, segmentation, and analysis. It is based on the concept of a topographic map, where the grey level values of an image represent elevations, and the image is viewed as a surface. The basic idea behind the watershed transform is to use the topographic map of an image to identify regions or objects within the image. This is done by ``flooding" the topographic map from its minima, which correspond to local minima in the image. As the flooding proceeds, the water from different minima starts to meet and form ridges or boundaries, which separate different regions in the image. More details about this method can be found for instance in \cite{meyer1}.

\begin{figure}
\begin{tabular}{ll}
\includegraphics[scale=0.83]{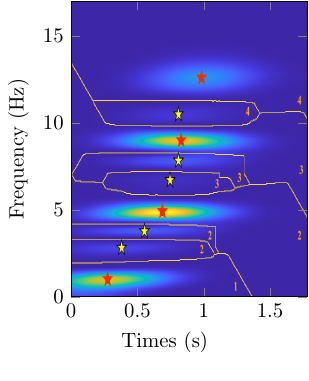}& 
\includegraphics[scale=0.83]{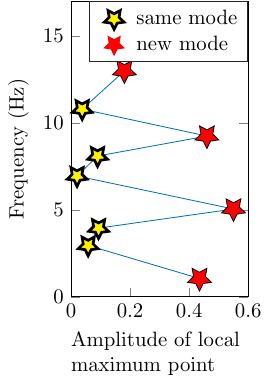} \vspace{-0.5cm}\\ (a) & (b) \end{tabular}
\caption{\label{ampwarp} (a): A warped spectrogram superposed with associated drainage basins and local maximum points (yellow and red stars). (b): Amplitude of each local maximum point sorted by descending frequencies; the top red star is assocated with mode number $4$, and so is the yellow star below it. The second red star from the top is associated to the mode number $3$, as well as the two following yellow stars, and so on.
}
\end{figure}

Using the shape of dispersion curves in Pekeris waveguides (see for instance Figure \ref{wrap_spectro}), we know that most of the energy of each mode is concentrated near its high frequencies. Therefore, we choose to apply the following algorithm to sort drainage basins: 
\begin{itemize}
\item We sort local maximum points $(t_i,\omega_i)$ associated with basins~$B_i$ by descending frequencies. 
\item We associate $B_1$ with the mode number $N$. For $i\geq 1$, if $S(t_{i+1},\omega_{i+1})<S(t_i,\omega_i)$, then we keep the same mode number, otherwise, we associate $B_{i+1}$ to a new mode number (the mode number of $B_i$ minus one). 
\end{itemize}
An example is shown in Figure \ref{ampwarp}. 

We still need to choose a good $t_0$ to separate modes. To do so, we maximize a quality factor defined to favor warpings where each mode is separated from the next one, both in amplitude and frequency distance. We privilege situations where each mode contains different drainage basins with an amplitude gap with the next mode. For a mode number $n$ associated to basins $i=I, \ldots ,J$, we define 
\begin{equation}\label{quality}
Q^{(n)}(t_0)=\left\{\begin{array}{l} 0 \quad \text{ if } I=J, \\
S(t_{J+1},\omega_{J+1})-S(t_{J},\omega_J) \quad\text{ else}. \end{array}\right. 
\end{equation}

\begin{figure}
\includegraphics[width=0.45\textwidth]{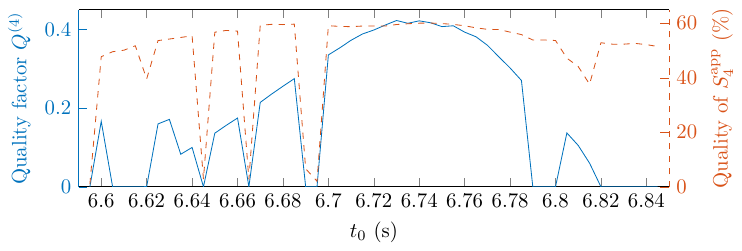}
\caption{\label{quality_comp} Comparison between the quality factor $Q^{(4)}$ and the quality of the reconstruction of $S_4$, with respect to $t_0$. The quality of the precision is computed using the formula $1-\Vert S_4-S_4^\app\Vert_2/\Vert S_4\Vert_2$, where $S_4$ is the true spectrogram of the mode $4$, $S_4^\app$ is the one obtained after modal separation, and $\Vert \cdot \Vert_2$ is the $L^2$-norm.}
\end{figure}

This quality factor is relatively simple and it encapsulates the warping quality very well as shown in Figure~\ref{quality_comp}. Its maximization gives the best $t_0$ to separate modes. 
If $Q^{(n)}$ vanishes everywhere, we remove the first condition and only measure the amplitude gaps between the basins. 

After choosing the best $t_0$ to maximize the quality factor $Q^{(N)}$, we remove all the drainage basins associated with the mode $N$, and we do the same process for the next mode $N-1$ with the quality factor $Q^{(N-1)}$ until the spectrogram is empty. The fully automated algorithm is summarized in Algorithm \ref{alg_filtering}.\\

\begin{algorithm}[ht]
\caption{Filtering of modal components in the received signal}\label{alg_filtering}
\hspace*{\algorithmicindent} \textbf{Input}: Signal $u$ measured on the regular grid \\ 
\hspace*{\algorithmicindent}\hspace*{\algorithmicindent} $\tau=(\tau_1,\ldots ,\tau_T)$ \\ 
\hspace*{\algorithmicindent}\hspace*{\algorithmicindent}\hspace*{\algorithmicindent}\hspace{5mm}  Number of propagative modes $N$ \\ 
    \hspace*{\algorithmicindent} \textbf{Output}: Modal components $u_1,u_2, \ldots, u_N$ 
\begin{algorithmic}[25]
\FOR{$n= N, \ldots, 2$}
\FOR{$\ell=1,\ldots , T$}
\STATE $t_0=\tau_{\ell}$
\STATE $\widetilde{u}\gets$ warping of $u$ associated to $\psi(t)=\sqrt{t^2+t_0^2}$ (see \eqref{eqwarp})
\STATE	$\widetilde{S} \gets$ spectrogram of $\widetilde{u}$ computed with the smallest possible value of $\sigma$ (see \eqref{def_spectro} and \eqref{h})
\STATE	$B_i,(t_i,\omega_i)\gets$  drainage basins of $\widetilde{S}$, local maxima of $\widetilde{S}$, sorted by descending frequencies 
\STATE	$Q^{(n)} (\ell)\gets$ quality factor at $t_0$ computed using $(B_i,t_i,\omega_i)$ (see \eqref{quality})
\ENDFOR
\STATE $t_0\gets$ argmax ($Q^{(n)}$)
\STATE	$B_i,(t_i,\omega_i)\gets$ drainage basins of $\widetilde{S}$ at $t_0$, local maxima of $\widetilde{S}$ at $t_0$
\STATE	mask $\gets$ $0\times \widetilde{S}$
\WHILE{$\widetilde{S}(t_{i+1},\omega_{i+1})<\widetilde{S}(t_i,\omega_i)$}
\STATE mask$=$mask$+\textbf{1}_{B_i}$ (indicator function of $B_i$)
\ENDWHILE
\STATE $u_n\gets$ inverse warping of the inverse spectrogram of $\widetilde{S}\times$mask 
\STATE $u\gets u-u_n$
\ENDFOR
\STATE $u_1\gets u$
\vspace{2mm}
\end{algorithmic}
\end{algorithm}

\begin{figure*}
\begin{center}
\begin{tabular}{lllll}
\includegraphics[width=0.19\textwidth]{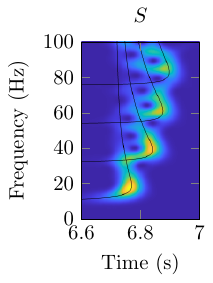} &
\includegraphics[width=0.19\textwidth]{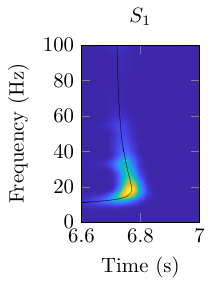} &\includegraphics[width=0.19\textwidth]{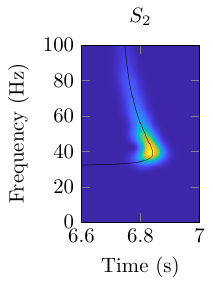} &\includegraphics[width=0.19\textwidth]{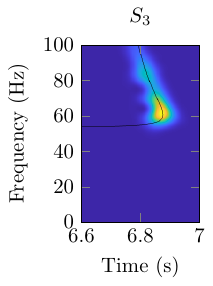} &\includegraphics[width=0.19\textwidth]{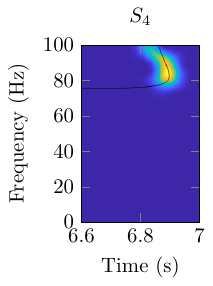} \vspace{-0.5cm}\\ (a) & (b) & (c) & (d) & (e)  
\end{tabular}\end{center}\vspace{-0.5cm}
\caption{\label{modal_separation} Initial spectrogram and separate modal components computed with $\sigma=20$ Hz and obtained with the warping algorithm on synthetically generated data in the Pekeris model given in Table \ref{tableau}-line 1. (a) initial spectrogram superposed with the dispersion curves. (b) spectrogram of mode 1. (c) spectrogram of mode 2. (d) spectogram of mode 3. (e) spectrogram of mode 4.}\vspace{-0.5cm}
\end{figure*}

\begin{figure*}\begin{center}
\begin{tabular}{lllll}
\includegraphics[width=0.19\textwidth]{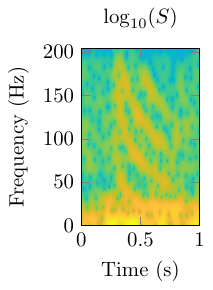} &
\includegraphics[width=0.19\textwidth]{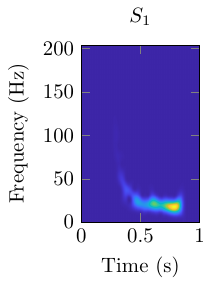} &\includegraphics[width=0.19\textwidth]{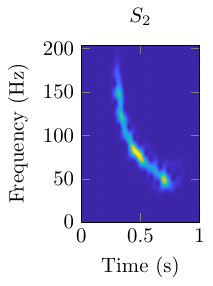} &\includegraphics[width=0.19\textwidth]{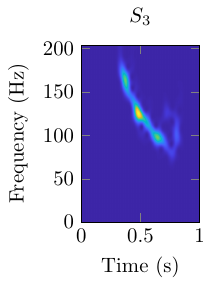} &\includegraphics[width=0.19\textwidth]{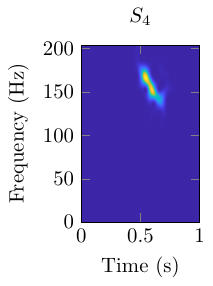} \vspace{-0.5cm}\\ (a) & (b) & (c) & (d) & (e)  
\end{tabular}\end{center}\vspace{-0.5cm}
\caption{\label{modal_separation_impulsive} Initial spectrogram and separate modal components computed with $\sigma=41$ Hz and obtained with the warping algorithm on data of a right whale gunshot \cite{thode1}. (a) initial spectrogram (in log scale). (b) spectrogram of mode 1. (c) spectrogram of mode 2. (d) spectogram of mode 3. (e) spectrogram of mode 4.}\vspace{-0.5cm}
\end{figure*}

\begin{figure*}\begin{center}
\begin{tabular}{lllll}
\includegraphics[width=0.19\textwidth]{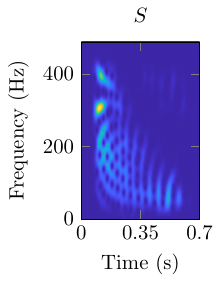} &
\includegraphics[width=0.19\textwidth]{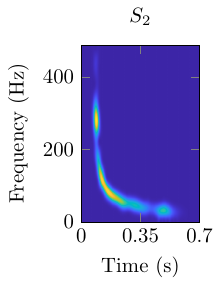} &\includegraphics[width=0.19\textwidth]{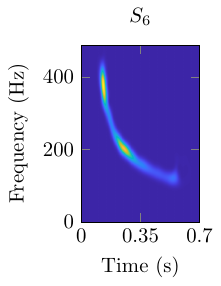} &\includegraphics[width=0.19\textwidth]{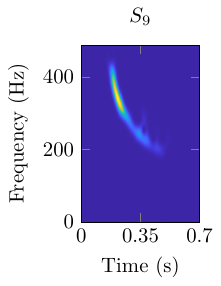} &\includegraphics[width=0.19\textwidth]{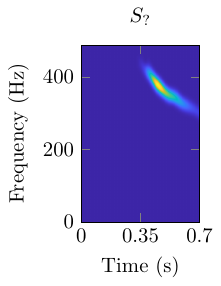} \vspace{-0.5cm}\\ (a) & (b) & (c) & (d) & (e)  
\end{tabular}\end{center}\vspace{-0.5cm}
\caption{\label{modal_separation_css} Initial spectrogram and separate modal components computed with $\sigma=98$ Hz and obtained with the warping algorithm on data of a combustive sound source \cite{bonnel2}. As in \cite{bonnel2}, we recover the 11 first modes, but due to a gap in the data, we cannot be sure of the mode numbers after that. We recover a total of 14 modal components using our warping algorithm. (a) initial spectrogram. (b) spectrogram of mode 2. (c) spectrogram of mode 6. (d) spectrogram of mode 9. (e) spectrogram of an unidentified mode. }\vspace{-0.5cm}
\end{figure*}

A modal separation done with this algorithm on synthetic data in the Pekeris model is presented in Figure~\ref{modal_separation}. Then, this algorithm is applied to experimental data with a right whale gunshot (Figure \ref{modal_separation_impulsive}) and a combustive sound source (Figure \ref{modal_separation_css}) and produces great reconstruction comparable to the ones proposed in \cite{bonnel1,bonnel2,thode1} but in a completely automated way. 

From now on, we assume that we can separate modal components and get good approximations of each modal spectrogram $S_n(t,\omega)$. In the following section, we will discuss the best action to use these spectrograms to get approximations of dispersion curves $\omega\mapsto t_n(\omega)$. 

\section{Reconstruction of the modal travel times}
\label{sec:reconstruction}%
In this section, for a fixed $n$, we aim at approaching the dispersion curves $\omega\mapsto t_n(\omega)$ given the modal spectrogram $S_n(t,\omega)$, and we consider a signal of the form
\begin{equation}
u(t)= u_n(t) +W(t) ,\quad u_n(t)=\frac{1}{2\pi} \int \hat{u}_n(\omega) e^{ i\omega t} d\omega,
\end{equation}
where $W(t)$ is a Gaussian additive noise satisfying $\E(W(t))=0$ and 
\begin{equation}
\CV(W(t),W(t'))=\delta^2 \exp\Big(-\frac{(t-t')^2}{2T_\delta^2}\Big). \end{equation}
This noise model seems to be a good approximation of the measurement noise and the environment noise \cite{aparicio1}, and can also account for the possible small errors in the previous separation method. 
As mentioned in Section \ref{sec:general}, we have a link between the spectrogram $S(t,\omega)$ and the dispersion curves $\omega\mapsto t_n(\omega)$ since 
\begin{equation}
S(t,\omega)\approx S^\app(t,\omega):=|A_n(\omega)|^2h(t-t_n(\omega))^2.
\end{equation}
Two competing methods would then allow to recover $t_n(\omega)$ using $S(t,\omega)$:
\begin{itemize}
\item Maximum method: we notice that 
\begin{align}
\nonumber \partial_t \sqrt{S^\app(t,\omega)}=0 \quad &\Leftrightarrow \quad \partial_t h(t-t_n(\omega))=0\\ \quad & \Leftrightarrow\quad t=t_n(\omega). 
\end{align}
To recover an approximated value of $t_n(\omega)$, we look for the maximum value of $t\mapsto S^\app(t,\omega)$ by solving the equation $\partial_t \sqrt{S^\app(t,\omega)}=0$. 
\item Mean method: since $h$ is a centered window function, 
\begin{align}
\nonumber \frac{\int_\R tS^\app(t,\omega)\dd t}{\int_\R S^\app(t,\omega)\dd t } &=\frac{\int_\R t h(t)^2\dd t+ t_n(\omega)\int_\R h(t)^2\dd t }{\int_\R h(t)^2\dd t } \\ &=t_n(\omega).
\end{align}
To recover an approximated value of $t_n(\omega)$, we can compute the quantity
\begin{equation} \frac{\int_\R t\phi(t)S^\app(t,\omega)\dd t}{\int_\R \phi(t)S^\app(t,\omega)\dd t },\qquad \phi(t)=
\exp\Big( -\frac{t^2}{2T_w^2}\Big),
\end{equation}
where $\phi$ is a weight function to account for the fact that we do not have access to measurements for every $t\in \R$.
\end{itemize}
To discriminate these two methods, we study the stability of each method and the reconstruction error given the noise level. In each case, we present the reconstruction error for a fixed window width $\sigma$, and then we optimize the width $\sigma$ to minimize the reconstruction error. 

\subsection{Maximum method}

First, we investigate the stability of the maximum method. In the following, for each function $\zeta_n$ depending on the mode $n$, we denote 
\begin{equation}
\Vert \zeta_n\Vert_\infty=\sup_{\tilde\omega \in (\omega_{c,n},+\infty)} |\zeta_n(\tilde \omega)|,
\end{equation}
where $\omega_{c,n}$ is the cut-off frequency of the mode $n$ \cite[§2.4.5.1]{jensen1}. 

\begin{prop}\label{max}
Let $t_n^\app(\omega)$ be an approximation of $t_n(\omega)$ defined as the unique solution of the equation $\partial_t \sqrt{S_n(t,\omega)}=0$. 
Defining 
$\sigma_{\rm lim}(\omega)=\omega_{c,n}/3-\omega/4$
(which do not depend on $\delta$, $T_\delta$, $\sigma$), we have
\begin{multline}
\E(|t_n^\app(\omega)-t_n(\omega)|) \\ \leq 
\frac{
\sup_{\tilde\omega \in (4\omega_{c,n}/3,+\infty)}|A_n'(\tilde\omega)|}{|A_n(\omega)|}+\frac{\Vert A_n'\Vert_\infty }{|A_n(\omega)|}\textbf{1}_{\sigma>\sigma_{\text{lim}}(\omega)}\\
+C_2\sigma\frac{\Vert A_n'\Vert_\infty \Vert\Phi_n''\Vert_\infty}{|A_n(\omega)|}+C_3\frac{\delta T_\delta^{1/2}}{|A_n(\omega)|\sigma^{3/2}} .
\end{multline}
This quantity is minimal when $\sigma=\sigma_{\text{opt}}$, where 
\begin{equation}\label{maxsopt}
\sigma_{\text{opt}}(\omega)\in \left\{\sigma_{\text{lim}}(\omega),\left(\frac{3\delta T_\delta^{1/2}2^{-11/4}}{\Vert A_n \Vert_\infty \Vert \Phi_n'' \Vert_\infty} \right)^{2/5}\right\}.
\end{equation}
Here, all  dimensionless quantities $C_i$ are explicit constants detailed in supplementary. 
\end{prop}

\begin{figure}
\begin{tabular}{ll}
\includegraphics[scale=0.9]{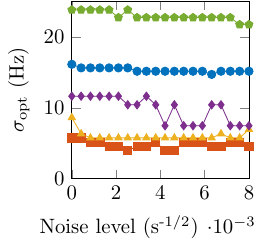} & 
\includegraphics[scale=0.9]{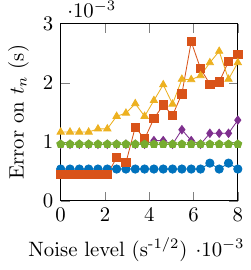} \vspace{-0.4cm} \\ (a) & (b) \end{tabular} 
\begin{center} \includegraphics{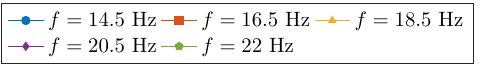} \end{center}
\caption{\label{comp_max} (a) Optimal choice of $\sigma$ to minimize the error of reconstruction of $t_n$ for increasing values of the noise level and different frequencies $f$. The noise level is defined as $\delta T_\delta^{1/2}/|A_n(\omega)|$. (b) Mean reconstruction error $\E(|t_n^\app(\omega)-t_n(\omega)|)$ for the optimal choice $\sigma=\sigma_{\text{opt}}$ with respect to the noise level $\delta T_\delta^{1/2}/|A_n(\omega)|$. The mean is computed using $50$ simulations, and $T_\delta=0.01s$. }
\end{figure}

Due to the term $\textbf{1}_{\sigma>\sigma_{\text{lim}}(\omega)}$ in the control of the reconstruction error, the optimal value of $\sigma$ cannot be computed easily and one needs to test the two values in \eqref{maxsopt} and determine each time which one provides the minimal error. However, in the Pekeris model, the most common situation seems to be the one where $\sigma_{\text{opt}}=\sigma_{\text{lim}}$, as illustrated in Figure \ref{comp_max} where we compute the optimal choice of $\sigma$ for different values of $\delta T_\delta^{1/2}$ and $\omega$. We also compute the associated reconstruction error $\E(|t_n^\app(\omega)-t_n(\omega)|)$.

\subsection{Mean method}
We now investigate the stability of the mean method.

\begin{prop}\label{mean}
Let $t_n^\app(\omega)$ be an approximation of $t_n(\omega)$ defined by
\begin{equation}t_n^\app(\omega)= \frac{\int_\R t\phi(t)S(t,\omega)\dd t}{\int_\R \phi(t)S(t,\omega)\dd t },\,\, \phi(t)=\exp\Big( -\frac{t^2}{2T_w^2}\Big). 
\end{equation}
If $\sigma t_n(\omega)>1$ and $\delta T_\delta^{1/2}/(t_n(\omega)^{1/2}|A_n(\omega)|)\ll 1$ then
\begin{multline}\label{contr_mean_1}
\E(|t_n^\app(\omega)-t_n(\omega)|)\leq \frac{\Vert A_n'\Vert_\infty \Vert A_n \Vert_\infty}{4|A_n(\omega)|^2}+D_1\frac{t_n(\omega)^2}{T_w}\\ +D_2\frac{\Vert A_n'\Vert_\infty^2t_n(\omega)}{|A_n(\omega)|^2}\sigma^2+D_3t_n(\omega)\frac{\delta T_\delta^{1/2}}{\sqrt{\sigma}|A_n(\omega)|}.
\end{multline}
This quantity is minimal for 
\begin{equation}\label{Di_2}
\sigma_{\text{opt}}(\omega)=D_4\frac{|A_n(\omega)|^{4/5}}{\Vert A_n' \Vert_\infty^{4/5}}\left(\frac{\delta T_\delta^{1/2}}{|A_n(\omega)|}\right)^{2/5},
\end{equation}
and provides an error 
\begin{multline}\label{Di_3} \E(|t_n^\app(\omega)-t_n(\omega)|)\leq \frac{\Vert A_n'\Vert_\infty \Vert A_n \Vert_\infty}{4|A_n(\omega)|^2}+D_1\frac{t_n(\omega)^2}{T_w}\\+D_5 \frac{t_n(\omega)\Vert A_n'\Vert_\infty^{2/5}}{|A_n(\omega)|^{2/5}}\left(\frac{\delta T_\delta^{1/2}}{|A_n(\omega)|}\right)^{4/5}.\end{multline} 
Here, all dimensionless quantities $D_i$ are explicit and depend on $\omega$, $T_w$, $A_n$ and $\Phi_n$. Their expressions are given in supplementary. 
\end{prop}

\begin{figure}\begin{tabular}{ll}
\includegraphics[scale=0.9]{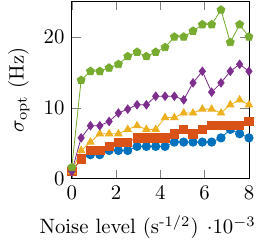} & 
\includegraphics[scale=0.9]{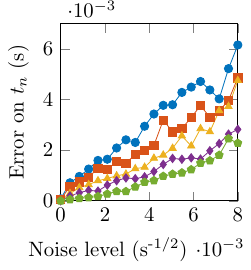} \vspace{-0.4cm} \\ (a) & (b) \end{tabular} 
\begin{center} \includegraphics{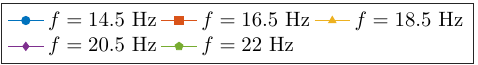} \end{center}
\caption{\label{comp_moyenne} (a) Optimal choice of $\sigma$ to minimize the error of reconstruction of $t_n$ for increasing values of the noise level and different frequencies $f$. The noise level is defined as $\delta T_\delta^{1/2}/|A_n(\omega)|$. (b) Mean reconstruction error $\E(|t_n^\app(\omega)-t_n(\omega)|)$ for the optimal choice $\sigma=\sigma_{\text{opt}}$ with respect to the noise level $\delta T_\delta^{1/2}/|A_n(\omega)|$. The mean is computed using $50$ simulations, and $T_\delta=0.01s$.}
\end{figure}

We illustrate these results in Figure \ref{comp_moyenne} where we plot $\sigma_{\text{opt}}(\omega)$ with respect to $\delta T_\delta^{1/2}/|A_n(\omega)|$. Numerically, we notice that for each plotted frequency, we have 
\begin{equation} \sigma_{\text{opt}}(\omega)=\mathcal{O}\left(\left(\frac{\delta T_\delta^{1/2}}{|A_n(\omega)|}\right)^{0.38}\right), \end{equation}
\begin{equation} \E(|t_n^\app(\omega)-t_n(\omega)|)=\mathcal{O}\left(\left(\frac{\delta T_\delta^{1/2}}{|A_n(\omega)|}\right)^{0.73}\right),\end{equation} 
which corroborates the theoretical results of Proposition~\ref{mean}.

\subsection{Comparison between both methods}

We can now compare the maximum and the mean methods using Propositions \ref{max} and \ref{mean}. Let us denote by $e_{\max}$ the error of reconstruction with the maximum method and by $e_{\text{mean}}$ the error with the mean method. First, we notice that, when there is no noise in the data (i.e., when $\delta T_\delta^{1/2} = 0$), and assuming that $T_w$ is large enough, we have 
\begin{equation}
e_{\max}\leq \frac{\Vert A_n'\Vert_{\infty}}{|A_n(\omega)|}, \qquad e_{\text{mean}}\leq \frac{\Vert A_n'\Vert_\infty \Vert A_n \Vert_\infty}{4|A_n(\omega)|^2}. 
\end{equation}
When $|A_n(\omega)|\approx \Vert A_n\Vert_\infty$ and without noise, the mean method turns out to be more precise. 

\begin{figure*}\begin{center}\begin{tabular}{lllll}
\includegraphics[width=0.19\textwidth]{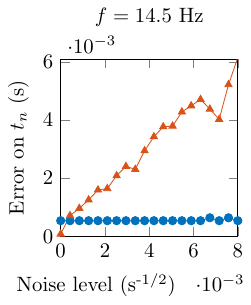} & 
\includegraphics[width=0.19\textwidth]{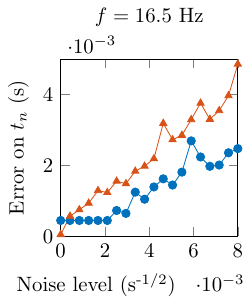} & 
\includegraphics[width=0.19\textwidth]{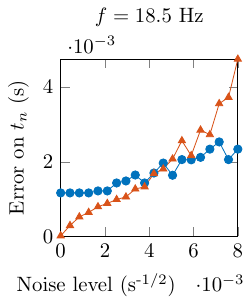} & 
\includegraphics[width=0.19\textwidth]{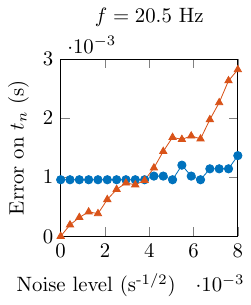} & 
\includegraphics[width=0.19\textwidth]{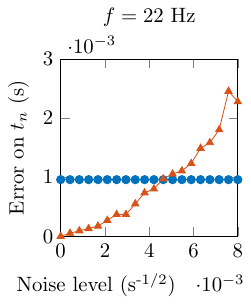} \vspace{-0.3cm} \\ (a) & (b) & (c) & (d) & (e) \end{tabular}
\includegraphics{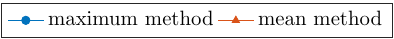}\end{center}\vspace{-0.5cm}
\caption{\label{comparison} Reconstruction error $\E(|t_n(\omega)-t_n^\app(\omega)|)(\sigma_{\text{opt}})$ with respect to the noise level $\delta T_\delta^{1/2}/|A_n(\omega)|$ for different frequencies. (a) $f=14.5$ Hz (b) $f=16.5$ Hz (c) $f=18.5$ Hz (d) $f=20.5$ Hz (e) $f=22$ Hz. The mean is computed with 50 simulations, and $T_\delta=0.01s$}\vspace{-0.5cm}
\end{figure*}

Assuming that $\sigma_{\text{opt}}=\sigma_{\text{lim}}$ for the maximum method when data are noisy, we notice that $e_{\max}$ grows as $\delta T_\delta^{1/2}/|A_n(\omega)|$ while $e_{\text{mean}}$ increases as $(\delta T_\delta^{1/2}/|A_n(\omega)|)^{4/5}$, meaning that the maximum method becomes more precise as the noise level grows. Numerical tests for different frequencies are presented in Figure~\ref{comparison} and confirm this comparison. We give the relative reconstruction error at the optimal choice $\sigma=\sigma_\text{opt}$ with the maximum and the mean methods for different frequencies in each case. 

\begin{figure}
\begin{tabular}{ll} 
\includegraphics[width=0.23\textwidth]{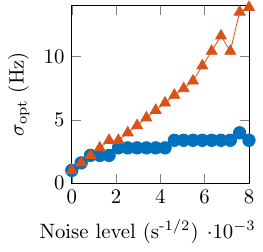} & 
\includegraphics[width=0.23\textwidth]{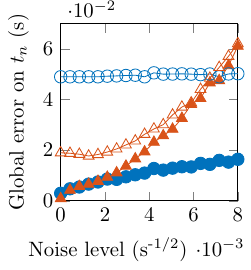} \vspace{-0.5cm} \\ (a) & (b) \end{tabular} 
\begin{center} \includegraphics{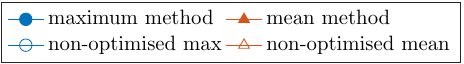}\end{center}
\caption{\label{comp_tot} (a) Optimal choice of $\sigma$ to minimize the global error in the reconstruction of $t_n$ for increasing values of the noise level $\delta T_\delta^{1/2}/|A_n(\omega)|$. (b) Global reconstruction error for the optimal choice $\sigma=\sigma_{\text{opt}}$, and comparisons with non-optimized errors with $\sigma=20$ Hz (choice in \cite{bonnel1}). The choice of threshold is $p=0.5$.}
\end{figure}

We then investigate the behavior of the global error 
\begin{equation}
\max_{\omega \in \Omega_n} \E(|t_n^\app(\omega)-t_n(\omega)|),
\end{equation}
where $\Omega_n$ is the set of frequencies
\begin{equation}
\label{eq:defOmegan}
\Omega_n = \left\{ \omega >0 \mbox{ s.t. } S_n(t_n^\app(\omega),\omega)>p \max(S) \right\} ,
\end{equation}
for a fixed threshold $p$ (its choice is discussed in the next section). This definition is very similar to the one presented in \cite{bonnel1}, where one must select the significative parts of the dispersion curves $\omega\mapsto t_n^\app(\omega)$ manually. We plot in Figure \ref{comp_tot} the optimal value $\sigma_\text{opt}$ to minimize the global error and the resulting global error for both methods.  Again, the mean method works better for a small noise level and the maximum method for a higher noise level. We also notice that the optimal value $\sigma_{\text{opt}}$ in the maximum method is very stable with respect to $\delta T_\delta^{1/2}/|A_n(\omega)|$, which makes it easy to calibrate  a priori.

Given the theoretical and numerical results presented in this section, we choose to work with the optimized maximum method to recover an approximated value of the dispersion curves. This choice differs from the one used in paper \cite{bonnel1} where the value $\sigma=20$ Hz is always chosen, and the mean method is used to compute the approximation of $t_n$. These non-optimized choices are represented in Figure \ref{comp_tot} for comparison purposes.

\begin{figure}\begin{tabular}{ll} 
\includegraphics[width=0.23\textwidth]{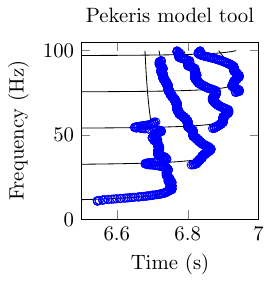} & 
\includegraphics[width=0.23\textwidth]{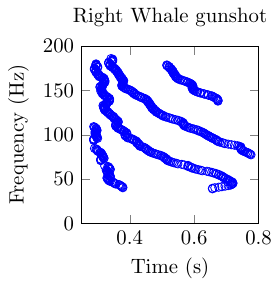} \vspace{-0.5cm}\\ (a) & (b) \\  
\raisebox{-2cm}{\includegraphics[width=0.23\textwidth]{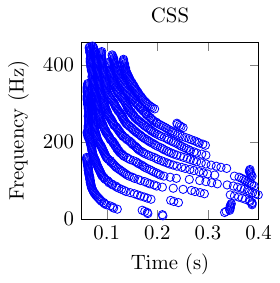}} & \hspace{1.1cm}\includegraphics{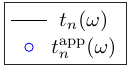} \vspace{-0.5cm} \\(c) & \end{tabular}
\caption{\label{tnapp} Approximated dispersion curves $\omega\mapsto t_n^\app(\omega)$ obtained using the maximum method with $\sigma=5f_{\max}/100$. We only present here the representative portions of curves such that $S_n(t_n^\app(\omega),\omega)>0.05 \max(S)$. (a) Pekeris model defined in Table \ref{tableau}-line 1 whose modal component have been extracted in Figure \ref{modal_separation}. (b) impulsive right whale gunshot  whose modal component have been extracted in Figure \ref{modal_separation_impulsive}. (c) combustive sound source  whose modal component have been extracted in Figure \ref{modal_separation_css}.
}
\end{figure}

Using the maximum method with an optimized choice $\sigma_{\text{opt}}=5f_{\max}/100$, we present in Figure \ref{tnapp} the approximated reconstructions $t_n^\app(\omega)$ obtained using Algorithm \ref{alg_tn} and the separation of modes presented in Algorithm \ref{alg_filtering}.

\begin{algorithm}[ht]
\caption{Reconstruction of dispersion curves}\label{alg_tn}
\hspace*{\algorithmicindent} \textbf{Input}: Modal component $u_n$ sampled at\\ 
\hspace*{\algorithmicindent}\hspace*{\algorithmicindent} frequency $f_{\max}$\\ 
\hspace*{\algorithmicindent}\hspace*{\algorithmicindent}\hspace*{\algorithmicindent}\hspace{5mm}  Set of frequencies $F$ used to compute the \\ 
\hspace*{\algorithmicindent}\hspace*{\algorithmicindent} spectrogram  \\ 
\hspace*{\algorithmicindent}\hspace*{\algorithmicindent}\hspace*{\algorithmicindent}\hspace{5mm}   Threshold $P=p\max(S)$  \\ 
    \hspace*{\algorithmicindent} \textbf{Output}: Dispersion curves $F\mapsto t_n^\app(F)$
\begin{algorithmic}[25]
\STATE $S_n\gets$ spectrogram of $u_n$ computed with $\sigma=5f_{\max}/100$ (see \eqref{def_spectro} and \eqref{h})
\FOR{$f\in F$}
\STATE $t_n^\app(f)=$argmax $S_n(f,\cdot)$
\IF{$S_n(t_n^\app(f),f)<P$ }
\STATE$t_n^\app(f)=$Nan \Comment{One could also remove the values from the list}
\ENDIF
\ENDFOR
\vspace{2mm}
\end{algorithmic}
\end{algorithm} 

\section{Inverse problem}
\label{sec:inverse}%
We now work with approximated values of the dispersion curves $\omega\mapsto t_n^\app(\omega)$ obtained using the method described in the previous section, and aim at recovering the range $r$ and the parameters of the Pekeris waveguide. Dispersion curves $\omega \mapsto t_n(\omega)$ do not have an explicit expression, but it is possible to compute excellent approximations by looking at the zeros of the relation \eqref{disp}. As described in \cite{bonnel1}, the most naive method to recover all the parameters would be to minimize the functional 
\begin{multline}\label{J}
J(r,c_{\rm w},c_{\rm b},\rho_{\rm w},\rho_{\rm b},D,dt) \\ =\Vert t_n(r,c_{\rm w},c_{\rm b},\rho_{\rm w},\rho_{\rm b},D,\cdot)-dt-t_n^\app(\cdot)\Vert^2_{\ell^2},
\end{multline}
where 
\begin{equation}
\Vert t_n \Vert^2_{\ell^2}= \sum_{n=1}^{N} \int_{\Omega_n} t_n(\omega)^2 {\rm d} \omega,\end{equation} 
and $\Omega_n$ is the set of frequencies associated to the mode~$n$ defined in \eqref{eq:defOmegan}.

However, different difficulties arise from this approach: 
\begin{itemize}
\item Firstly, all parts of curves $\omega\mapsto t_n^\app(\omega)$ may not be significant and one must choose an appropriate threshold $p$ to define $\Omega_n$ in \eqref{eq:defOmegan}. Especially we prove in the proofs of Propositions \ref{max} and \ref{mean} that the reconstruction error increases as $1/|A_n(\omega)|$, which leads to large errors for frequencies with low energy.

\item Secondly, the functional $J$ is not convex and its optimization can be costly. 
\item Thirdly, the minimum in \eqref{J} may be reached at non-physical values, especially for $c_{\rm w}$ and $\rho_{\rm w}$ which actually do not vary much in shallow waters. 
\end{itemize}

To solve the first issue, we propose to determine empirically the optimal threshold value, denoted as $p$, by executing the algorithm on the Pekeris model. This will allow us to identify the ideal value of $p$ that minimizes the relative reconstruction error for all the parameters.

To solve the second and third issues, the paper \cite{bonnel1} chooses to reduce the number of variables and to fix the values of $c_{\rm w},\rho_{\rm w},\rho_{\rm b},D$. The optimization is done with the unknown $r,c_{\rm b}, dt$ and turns out to be much simpler. We develop a more powerful approach and adapt a Bayesian approach with prior knowledge of the values of some parameters. This approach helps us account for the fact that the Pekeris model is just an approximation of the actual propagation in shallow water. To do so, we add a penalization term to the functional $J$ depending on hyper-parameters accounting for our confidence in the a priori values $c_{\rm w}^0$, $\rho_{\rm w}^0$, $\rho_{\rm b}^0, D^0$ of the parameters $c_{\rm w},\rho_{\rm w},\rho_{\rm b}$ and $D$. 

These hyper-parameters are usually determined by training on big sets of data \cite{mohammad1,pascal1}. However, we only have two experiences in very different contexts. Despite this, we know that the estimates on the water layer are usually very reliable, while those on $\rho_{\rm b}$ and $D$ are more uncertain. For this reason, we propose the following choice of hyper-parameters for the optimization and define a new functional
\begin{multline}
\label{jtilde}
\widetilde{J}(r,c_{\rm w},c_{\rm b},\rho_{\rm w},\rho_{\rm b},D,dt)=J(r,c_{\rm w},c_{\rm b},\rho_{\rm w},\rho_{\rm b},D,dt) \\
+\alpha \Big(\Big|\frac{D-D^0}{D^0}\Big|^2+\Big|\frac{\rho_{\rm b}-\rho_{\rm b}^0}{\rho_{\rm b}^0}\Big|^2\\ +10\Big|\frac{\rho_{\rm w}-\rho_{\rm w}^0}{\rho_{\rm w}^0}\Big|^2+10\Big|\frac{c_{\rm w}-c_{\rm w}^0}{c_{\rm w}^0}\Big|^2\Big).
\end{multline}
Here, $\alpha$ is chosen so that the penalization part is of the same order as $J$. We will show in the following section that this choice of hyper-parameters yields good results for both experimental setups and the Pekeris model. By minimizing the penalized functional $\widetilde{J}$, we obtain the \textit{a posteriori} maximum of the corresponding Bayesian approach, which incorporates \textit{a priori} information on $c_{\rm w}^0$, $\rho_{\rm w}^0$, $\rho_{\rm b}^0$, and $D^0$. For a more detailed explanation, see \cite{gribonval1}.

First, we investigate the optimal choice of threshold $p$ on the Pekeris model. Then, we present the reconstruction obtained by our algorithm on experimental data. 

\subsection{Choice of the threshold $p$}

The choice of the threshold $p$ is important: if $p$ is too close to $1$, then the optimization is done with very few measurements and can become very sensitive to measurement errors. On the other hand, if $p$ is too close to $0$, then the reconstruction $t_n^\app$ can be very far from the actual values $t_n$ which can tamper the reconstruction of the parameters. The good choice results from a trade-off to keep enough measurement points to prevent errors and remove non-significative measurements. 
Using the dispersion curves presented in Figure \ref{tnapp} for the Pekeris model, we minimize the functional $\widetilde{J}$ for different choices of $p$ and present the relative reconstruction errors in Table \ref{min_pek}. 
Given these data, we choose from now on to work with the choice of threshold $p=0.4$ and provide the reconstruction Algorithm \ref{alg_rec}.

\begin{table}
\begin{tabular}{|c|ccccccc|}
\hline
$p$ & $r$ & $c_{\rm w}$ & $c_{\rm b}$ & $\rho_{\rm w}$ & $\rho_{\rm b}$ & $D$ & $dt$ \\ \hline 

0.2 & 2 \% & 2 \%&2 \% & $<$1\%& 2\%&3\% & 1\% \\\hline

 \rowcolor{gray!30} 0.4 &$<$1 \% &$<$1 \%&$<$1 \%&$<$1\% & 2\%&$<$1\% &$<$1\% \\\hline
0.6 &$<$1 \%&$<$1 \%&$<$1 \%&$<$1\%&3\%&2\% & 1\% \\\hline
0.8 &$<$1 \%&$<$1 \%&$<$1 \%&$<$1\%&9\%& 3\% & $<$1\% \\  \hline
\end{tabular}
\caption{\label{min_pek} Relative reconstruction errors of $r,c_{\rm w},c_{\rm b},D,\rho_{\rm w},\rho_{\rm b}$ in the Pekeris model with respect to the threshold $p$.}
\end{table}

\begin{algorithm}[ht]
\caption{Recovery of parameters}\label{alg_rec}
\hspace*{\algorithmicindent} \textbf{Input}: Measured signal $u$ at times $\tau={\tau_1,\ldots ,\tau_T}$ \\ 
\hspace*{\algorithmicindent}\hspace*{\algorithmicindent} sampled at frequency $f_{\max}$ \\ 
\hspace*{\algorithmicindent}\hspace*{\algorithmicindent}\hspace*{\algorithmicindent}\hspace{2mm}  Set of frequencies $F$ to compute the\\ 
\hspace*{\algorithmicindent}\hspace*{\algorithmicindent} spectrogram \\ 
\hspace*{\algorithmicindent}\hspace*{\algorithmicindent}\hspace*{\algorithmicindent}\hspace{2mm}  Number of modes $N$  \\ 
\hspace*{\algorithmicindent}\hspace*{\algorithmicindent}\hspace*{\algorithmicindent}\hspace{2mm}  Estimated values of $c_{\rm w}^0$, $\rho_{\rm w}^0$, $\rho_{\rm b}^0$, and $D^0$  \\ 
    \hspace*{\algorithmicindent} \textbf{Output}: Recovered parameters $r,c_{\rm w},c_{\rm b},D,\rho_{\rm w},\rho_{\rm b}$
\begin{algorithmic}[25]
\STATE $S\gets$ spectrogram of $u$ computed with $\sigma=5f_{\max}/100$ (see \eqref{def_spectro} and \eqref{h})
\STATE $u_1,\ldots, u_n\gets$ filtering of $u$ using Algorithm \ref{alg_filtering}
\STATE $P\gets 0.4\max(S)$ 
\STATE $t_1^\app\ldots, t_n^\app\gets$ result of Algorithm \ref{alg_tn} on $u_1, \ldots u_n$ with the threshold $P$
\STATE  $r^0,c_{\rm b}^0,dt^0\gets$ arbitrary values
\STATE $\alpha\gets J(r^0,c_{\rm w}^0,c_{\rm b}^0,\rho_{\rm w}^0,\rho_{\rm b}^0,D^0,dt^0)$
\STATE $(r,c_{\rm w},c_{\rm b},\rho_{\rm w},\rho_{\rm b},D,dt)\gets$ minimization of $\widetilde{J}$ defined in \eqref{jtilde} and starting at $(r^0,c_{\rm w}^0,c_{\rm b}^0,\rho_{\rm w}^0,\rho_{\rm b}^0,D^0,dt^0)$
\vspace{2mm}
\end{algorithmic}
\end{algorithm}

\subsection{Experimental data}

\begin{figure}\begin{tabular}{ll} 
\includegraphics[width=0.23\textwidth]{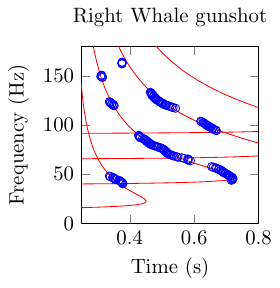} & 
\includegraphics[width=0.23\textwidth]{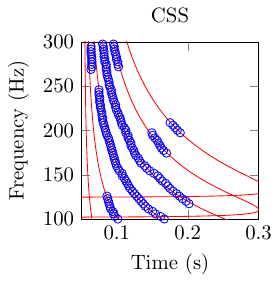} \vspace{-0.5cm} \\ (a) & (b) \end{tabular}
\begin{center} \includegraphics{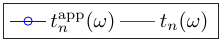}\end{center}\vspace{-0.5cm}
\caption{\label{resultat} Dispersion curves $\omega \mapsto t_n(\omega)$ obtained by minimizing the functional $\widetilde{J}$ on data $t_n^\app(\omega)$ presented in Figure \ref{tnapp}-(b-c).
(a) Right whale gunshot data. (b) Combustive sound source data.}
\end{figure}

We now minimize the functional $\widetilde{J}$ with a threshold $p=0.4$ on the approximated dispersion curves obtained in Figure \ref{tnapp} for the experimental data. As explained in \cite{bonnel1,bonnel2}, the environment of the combustive sound source experiment is only comparable to a Pekeris waveguide when frequencies are between $100$ Hz and $300$ Hz, and we only consider these data for the inversion. In Figure \ref{resultat}, we present the dispersion curves $t_n$ associated with the parameters minimizing the functional $\widetilde{J}$. For the impulsive right whale gunshot, we obtain the following parameters: 
$r=8.8\text{km}$, $c_{\rm b}=1727\text{m/s}$, $dt=5.9s$, 
$c_{\rm w}=1449\text{m/s}$, $\rho_{\rm w}=1007\text{kg/m}^3$, $\rho_{\rm b}=1481 \text{kg/m}^3$, $D=51.2\text{m}$. 
For the combustive sound source, we obtain $r=4.8\text{km}$, $c_{\rm b}=1649\text{m/s}$, $dt=3.2s$, $c_{\rm w}=1476\text{m/s}$, $\rho_{\rm w}=996\text{kg/m}^3$, $\rho_{\rm b}=1282 \text{kg/m}^3$, $D=73.1\text{m}$. 
These results are consistent with the ground truth and those exposed in Refs.~\cite{bonnel1,bonnel3,thode1} and in Table \ref{tableau}. We even find a better localization result in the combustive sound source case, thanks to the relaxation of the parameters $c_{\rm w},\rho_{\rm w}, \rho_{\rm b}$, and~$D$. 

\section{Conclusion}

In this paper, we have presented theoretical tools and numerical methods for automating the process of source localization and medium parameter estimation in shallow waters. Our approach involves 1) using a watershed algorithm and optimized warping to effectively separate modal components of the signal and 2) developing a stability-optimized maximum method to obtain accurate approximations of dispersion curves. 
The method also makes it possible to incorporate prior estimates on certain medium parameters. 
The application of this automated method on two experimental data sets show that it accurately recovers source location and additional medium parameters.


The definition of the quality factor of the warping in \eqref{quality} and the choices of hyper-parameters in \eqref{jtilde} follow from theoretical considerations. 
We have shown that these choices provide reliable estimates of the parameters of interest and that the automated method demonstrates stable performance with respect to the chosen values of the hyper-parameters. 
They could, however, be updated in the presence of additional information about the experimental setup. 
Indeed, with further experimental data, it is possible to refine the choices to improve the accuracy of our estimates even further.
To summarize, our current method demonstrates significant potential for completely automating the process of source localization and medium parameter estimation in shallow waters.
\footnote{See supplementary materials attached to find the mathematical developments of the paper, and at \url{https://github.com/niclas-angele/source_localization} for all the Matlab codes associated to the article.}

\bibliography{biblio}

\end{document}